\documentclass[aps,prd,preprint,tightenlines,groupedaddress,amsmath,amssymb,superscriptaddress]{revtex4}  
\usepackage{graphicx}

\begin{document}
\title{Gauging the Contribution of X-ray Sources to Reionization Through the Kinetic Sunyaev-Zel'dovich Effect}

\author{Eli Visbal}
\email[]{evisbal@fas.harvard.edu}
\affiliation{Jefferson Laboratory of Physics, Harvard University,
Cambridge, MA 02138} 
\affiliation{Institute for Theory $\&$ Computation, Harvard University,
60 Garden Street, Cambridge, MA 02138}

\author{Abraham Loeb}
\email[]{aloeb@cfa.harvard.edu}
\affiliation{Institute for Theory $\&$ Computation, Harvard University,
60 Garden Street, Cambridge, MA 02138}

\begin{abstract}
Measurements of the kinetic Sunyaev-Zel'dovich (kSZ) effect from
instruments such as the South Pole Telescope (SPT) and the Atacama
Cosmology Telescope (ACT) will soon put improved constraints on
reionization.  Popular models assume that UV photons alone are
responsible for reionization of the intergalactic medium.  We explore
the effects of a significant contribution of X-rays to reionization on
the kSZ signal.  Because X-rays have a large mean free path through
the neutral intergalactic medium, they introduce partial ionization in
between the sharp-edged bubbles created by UV photons.  This smooth
ionization component changes the power spectrum of the cosmic
microwave background (CMB) temperature anisotropies.  We quantify this
effect by running semi-numerical simulations of reionization.  We test
a number of different models of reionization without X-rays that have
varying physical parameters, but which are constrained to have similar
total optical depths to electron scattering.  These are then compared
to models with varying levels of contribution to reionization from
X-rays.  We find that models with more than a 10$\%$ contribution from
X-rays produce a significantly lower power spectrum of temperature
anisotropies than all the UV-only models tested.  The expected
sensitivity of SPT and ACT may be insufficient to distinguish between
our models, however, a non-detection of the kSZ signal from the epoch
of reionization could result from the contribution of X-rays.  It
will be important for future missions with improved sensitivity to
consider the impact of X-ray sources on reionization.

\end{abstract}

\maketitle

\section{Introduction}
The epoch of reionization (EoR) is an important milestone in the
history of our universe \cite{avibook}.  There has been much
theoretical work predicting the details of reionization, however many
of the predictions have yet to be verified observationally.  An
important probe of this process is the small scale temperature
fluctuations in the cosmic microwave background (CMB).  In particular,
secondary anisotropies are imprinted on top of the primordial CMB from
the Doppler shift associated with CMB photons scattering off free
electrons in between the observer and the surface of last scattering
at $z \approx 1100$.  This is known as the kinetic Sunyaev-Zel'dovich (kSZ)
effect.  Because only ionized regions can contribute to this effect,
the power spectrum of kSZ fluctuations is sensitive to details of the
reionization history
\cite{2005ApJ...630..643M,2005ApJ...630..657Z,2007ApJ...660..933I,1998ApJ...508..435G,1998PhRvL..81.2004K}.

The most popular models of reionization assume that the UV photons
from early stars drive the ionization of the intergalactic medium
(IGM).  Due to the short mean free path (MFP) of UV photons through
the neutral IGM, the resulting topology of reionization is a two phase
IGM consisting of ionized bubbles surrounded by regions of neutral
gas \cite{2001PhR...349..125B}.

Another possibility is that X-ray photons make a significant
contribution to reionization
(e.g. \cite{2004MNRAS.352..547R,2000ApJ...534...11H,2001ApJ...563....1V,2009MNRAS.396.1106W,2009ApJ...703.2113V}).
These X-rays could originate from accretion onto massive black holes
in galactic nuclei, X-ray binaries, or inverse Compton scattering in
supernova remnants \cite{2006PhR...433..181F}. Because the MFP of
X-rays through the IGM is much longer than UV photons this will result
in a more uniform background of ionizing photons.  Thus, a significant
contribution of X-rays can qualitatively change reionization by
creating a partially ionized IGM in between the bubbles produced by UV
photons.

In this paper we use semi-numerical simulations to gauge the imprint
 of X-ray sources during reionization on the kSZ signal.  Because of
 the smooth ionization caused by X-rays, the so called ``patchy''
 component of the kSZ power spectrum can be substantially reduced.
We find that models with more than a 10$\%$ contribution from
X-rays produce a significantly lower power spectrum of temperature
anisotropies than all the UV-only models tested. 


The paper is structured as follows.  In $\S$2, we describe our method
of estimating the kSZ signal including a description of our
semi-numerical simulations of reionization with and without a
contribution from X-rays.  In $\S$3, we report our results, which
include changes in the CMB power spectrum due to the contribution of
X-ray sources to reionization.  In $\S$4 we summarize our main
conclusions.  Throughout, we assume a $\Lambda$CDM cosmology with
$\Omega_\Lambda= 0.73$, $\Omega_m = 0.27$, $\Omega_b = 0.0456$, $h =
0.7$, and $\sigma_8 = 0.81$ \cite{2011ApJS..192...18K}.

\section{Method}
\subsection{The Kinetic SZ Effect}
The kSZ effect is the contribution to fluctuations in the CMB
temperature associated with the Doppler shift of CMB photons Thompson
scattered off free electrons moving at some radial velocity relative
to the observer.  The kSZ temperature fluctuation in a particular
direction is given by
\begin{equation}
\label{dTintegral}
\frac{\Delta T}{T} = \sigma_{\rm T} \int d\eta e^{-\tau(\eta)} a n_e v_{||}, 
\end{equation}
where $\sigma_{\rm T}$ is the Thomson cross section, $\tau(\eta)$ is
the optical depth from the observer to the position corresponding to
conformal time $\eta$, $a$ is the cosmic scale factor, $n_e$ is the
number density of free electrons, and $v_{||}$ is the component of the
velocity along the line of sight.

We wish to calculate the impact on the kSZ effect from an X-ray
contribution to reionization using semi-numerical simulations of
the EoR.  These simulations will examine varying levels of X-ray
contribution to reionization.  We generate a number of simulation
cubes that each span a solid angle of one square degree on the sky
(corresponding to a comoving length of $L \approx 150-200$ Mpc from
$z=6-20$) and have a resolution of $256^3$ separate volume elements.
In each 3D element of the simulation we calculate the matter
over-density, velocity, and ionization fraction.  Cubes are produced at
appropriate cosmic times and stacked along the line of sight to give
continuous coverage from a redshift of $z=6$ to $z=20$, corresponding
to the EoR.  We then perform the integral in Eq. (\ref{dTintegral})
along each line of sight to produce a 2D map of CMB temperature
fluctuations.

We compute the density and velocity fields of our simulations using
the 21cmFAST software package \cite{2011MNRAS.411..955M}.
Non-linearities are included using leading order perturbation theory
(Zel'dovich approximation).  We expect this to be a reasonable
approximation, as the power spectra of the density field using this
approach agree well with that from detailed hydrodynamic simulations
at the redshifts and scales relevant for our calculations
\cite{2011MNRAS.411..955M}.  We generate our ionization fields with
the model described below.  Note that we do not produce data cubes at
redshifts below $z=6$.  The kSZ fluctuations from these redshifts will
be independent of reionization history and we incorporate their effect
analytically.

\subsection{Reionization Model} 
To simulate the reionization history we use a semi-numerical approach
similar to that described in \cite{2009MNRAS.396.1106W}.  The basic
assumption in our prescription is that there is a fixed number of
ionizing photons, $N_{\rm ion}$, released into the IGM for each baryon
which is incorporated into galaxies.  From extended Press-Schechter
theory it is possible to calculate the rate at which matter will
collapse into dark matter halos \cite{1991ApJ...379..440B}.  In this
way one can calculate the ionized fraction in a region of a given size
and density analytically.  For the case without X-rays where all
ionizing photons are assumed to be UV, the evolution of the ionization
in a spherical region of comoving radius $R$ with over-density
$\delta$ relative to the cosmic mean is given by \cite{2007MNRAS.375.1034W}
\begin{equation}
\frac{dQ_{R,\delta}}{dt} = \frac{N_{\rm ion}}{0.76} \left ( Q_{R,\delta} \frac{dF(M_{\rm ion})}{dt} + (1-Q_{R,\delta}) \frac{dF(M_{\rm min})}{dt} \right ) - B,
\end{equation}
where the $0.76$ accounts for the cosmic mass fraction of hydrogen,
 $Q_{R,\delta}$ is the mass averaged ionized fraction of hydrogen, and
 $F(M)$ (which depends implicitly of $\delta$ and $R$) is the fraction
 of mass which has collapsed into dark matter halos above mass $M$,
\begin{equation}
\label{uvdiffeq}
F(M) = {\rm erfc} \left ( \frac{\delta_c(z) - \delta(z=0)}{\sqrt{2(\sigma(M)^2-\sigma(M_R)^2)}} \right ),
\end{equation}
where $\delta_c = 1.686/D(z)$ is the threshold for spherical collapse
at $z \gg 1$ extrapolated to redshift zero with linear theory, $D(z)$
is the linear growth factor of density fluctuations, $\delta(z=0)$ is
the linear over-density of the region at the current time, and the
$\sigma$'s are the root mean squared density fluctuations on scales
corresponding to mass $M$ and radius $R$ today. The minimum mass of
dark matter halos that can host galaxies is denoted by $M_{\rm min}$
in the neutral IGM and by $M_{\rm ion}$ in the ionized and
photo-heated IGM.  We parameterize this mass in terms of the virial
temperature which for $z \gg 1$ is given by
\begin{equation}
\label{tvir}
T_{\rm vir}=10^4 \left ( \frac{M}{10^8 M_\odot} \right )^{1/3} \left ( \frac{1+z}{10} \right ) {\rm K}.
\end{equation}
The recombination rate is given by
\begin{equation}
\label{UVrecomb}
B = \alpha_{\rm B} C n_{\rm H}^0 (1+z)^3 (1+\delta) Q_{R,\delta},
\end{equation}
where $\alpha_{\rm B}=2.6 \times 10^{-13} {\rm cm^3/s}$ is the case B
recombination coefficient for $T=10^4$ K \cite{Osterbrock}, $C=\langle n_{e}^2
\rangle/\langle n_{e} \rangle^2$ is the clumpiness of the ionized gas
in the IGM and $n_{\rm H}^0$ is the density of hydrogen gas today.


In order to produce the 3D simulation of the ionization fraction we
smooth a random realization of the linear density field in our
simulation in spheres of varying radius and solve Eq. (\ref{uvdiffeq})
as a function of the size and over-density at each point.  At this
stage, regions for which $Q_{R,\delta} > 1$ are taken to be bubbles
and the ionization fraction is set to one while other regions are
taken to be completely neutral.  Next, the equation is solved on the
scale of each spatial pixel and then used to determine the ionization
fraction in regions not encompassed by ionization bubbles.

We use a similar approach to calculate the ionization fraction when
X-rays make a significant contribution to reionization.  This approach
has been adapted from \cite{2007MNRAS.375.1034W}.  The key simplifying
assumption we make is that X-rays have a fixed comoving MFP and that
X-ray photons uniformly ionize regions of that size.  The comoving MFP
of X-rays of energy $E_X$ through a neutral medium is given by
\cite{2004ApJ...604..484M,2006PhR...433..181F},
\begin{equation}
\lambda \approx 180 \left( \frac{1+z}{15}\right)^{-2} \left( \frac{E_X}{1{\rm keV}}\right)^3 {\rm Mpc}.
\end{equation}
We assume a constant MFP of $\lambda=20$ Mpc, but find that the
measured angular power spectrum of temperature anisotropies is
essentially unchanged if we set $\lambda=50$ Mpc.  UV photons have a
very short MFP compared to the size of the spatial pixels in our
simulation.  Thus, when considering a region smaller than the MFP of
X-rays one must consider the X-ray sources from a larger surrounding
region with a radius equal to the X-ray MFP, $\lambda$, while the
ionization from UV photons will only depend on the local over-density.
We capture this behavior by solving a system of 6 coupled ordinary
differential equations (ODEs) which compute the ionization history
(including the relative amounts from X-rays versus UV photons) on
various scales.  As in the UV only case, we smooth the density field
on all scales to identify ionized bubbles, but in addition
simultaneously track the ionization history and production of X-ray
photons from surrounding regions with radius equal to the X-ray MFP
($\lambda$).  The system of 6 ODEs (adapted from
\cite{2007MNRAS.375.1034W}) is
\begin{equation}
\label{firstode}
\frac{dQ_R}{dt} = \frac{dQ_{R,UV}}{dt} + \frac{dQ_{R,X}}{dt}
\end{equation}

\begin{equation}
\frac{dQ_\lambda}{dt} = \frac{dQ_{\lambda,UV}}{dt} + \frac{dQ_{\lambda,X}}{dt}
\end{equation}

\begin{equation}
\frac{dQ_{R,UV}}{dt} = (1-X_{\rm frac})\frac{N_{\rm ion}}{0.76} \left ( Q_{R} \frac{dF(M_{\rm ion})}{dt} + (1-Q_{R}) \frac{dF(M_{\rm min})}{dt} \right ) - B_3
\end{equation}

\begin{equation}
\frac{dQ_{\lambda,UV}}{dt} = (1-X_{\rm frac})\frac{N_{\rm ion}}{0.76} \left ( Q_{\lambda} \frac{dF(M_{\rm ion})}{dt} + (1-Q_{\lambda}) \frac{dF(M_{\rm min})}{dt} \right ) - B_4
\end{equation}

\begin{equation}
\frac{dQ_{R,X}}{dt} = X_{\rm frac}\frac{N_{\rm ion}}{0.76} \left ( Q_{R} \frac{dF(M_{\rm ion})}{dt} + (1-Q_{R}) \frac{dF(M_{\rm min})}{dt} \right ) - B_5  \quad {\rm if} \quad  R > \lambda 
\end{equation}
\begin{equation}
\frac{dQ_{R,X}}{dt} = X_{\rm frac}\frac{N_{\rm ion}}{0.76} \left (\frac{1+\delta_\lambda}{1+\delta_R}  \right ) \left ( Q_{\lambda} \frac{dF(M_{\rm ion})}{dt} + (1-Q_{\lambda}) \frac{dF(M_{\rm min})}{dt} \right ) - B_5 \quad {\rm if} \quad R < \lambda  
\end{equation} 

\begin{equation}
\label{lastode}
\frac{dQ_{\lambda,X}}{dt} = X_{\rm frac}\frac{N_{\rm ion}}{0.76} \left ( Q_\lambda \frac{dF(M_{\rm ion})}{dt} + (1-Q_\lambda) \frac{dF(M_{\rm min})}{dt} \right ) - B_6,
\end{equation}
where $Q_R$ and $Q_\lambda$ are the mass averaged ionization fractions
on scales of $R$ and $\lambda$ respectively.  $Q_{R,UV}$ is the
ionization on scale $R$ due to UV photons and $Q_{R,X}$,
$Q_{\lambda,X}$, and $Q_{\lambda,UV}$ are the various ionization
fractions from either X-rays or UV photons ionization on the two
scales of interest.  The fraction of ionizing photons which are X-rays
is denoted by $X_{\rm frac}$.  Note that the time derivative of the
collapsed fraction in each equation is evaluated on the same scale as
the $Q$ value it is multiplied with.

The various $B_i$ terms are recombination rates.  In the UV case,
Eq. (\ref{UVrecomb}) was derived by assuming that all regions are either
completely ionized or neutral.  With a significant contribution from
X-rays this will not be the case.  We assume that X-rays uniformly
ionize the IGM on scale $\lambda$ and UV photons completely ionize
bubbles in the IGM around their sources.  In order to calculate the
recombination terms we calculate the volume of the IGM which is
uniformly ionized, $V_{\rm uniform}$, and the fraction in ionized bubbles,
$V_{\rm bub}$.  The uniform X-ray ionization also allows bubbles to
grow larger than they would from the UV photons alone; we denote this
additional ``fringe'' volume with $V_{\rm fringe}$.  These add up to
the total volume of the spherical regions we consider, $V$, and are
described by
\begin{equation}
V_{\rm bub}/V =  Q_{UV}
\end{equation}
\begin{equation}
V_{\rm fringe}/V = \frac{Q_{UV}Q_X}{1-Q_X}
\end{equation}
\begin{equation}
V_{\rm uniform} = V - V_{\rm bub} - V_{\rm fringe},
\end{equation}
where either the UV or X-ray ionization must be evaluated on the
relevant length scale.  It follows that various recombination rate
terms are given by
\begin{equation}
B_3 = \alpha_{\rm B} C n_{\rm H}^0 (1+z)^3 (1+\delta_R) \frac{V_{\rm bub}}{V},
\end{equation}
\begin{equation}
B_5 = \alpha_{\rm B} C n_{\rm H}^0 (1+z)^3 (1+\delta_R) \left (\frac{V_{\rm fringe}}{V} + Q_{R,X}^2\frac{V_{\rm uniform}}{V}\right ).
\end{equation}
$B_4$ and $B_6$ are the same as $B_3$ and $B_5$ except that the
appropriate values for $\delta_\lambda$ and the volumes corresponding
to scale $\lambda$ must be used.  Note that this treatment of the
recombination rates was not included in previous work.  We do not
include heating due to X-rays which could further suppress the
formation of low mass galaxies.  This would be straightforward to
incorporate in the future \cite{2007MNRAS.375.1034W}.

We smooth our density field on all scales and if $Q_{R} > 1$ the
region $R$ is set to be completely ionized.  After the entire cube has
been searched on all relevant scales we solve
Eqs. (\ref{firstode}-\ref{lastode}) on the scale of each spatial pixel
to determine the ionization fraction outside of the completely ionized
bubbles.

\section{Results}
\begin{figure}
\includegraphics{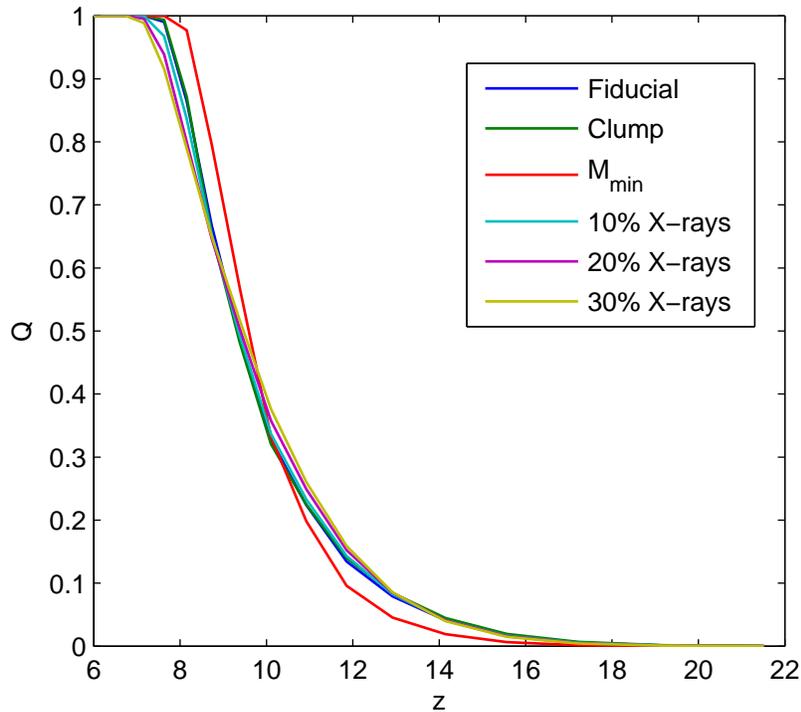}
\caption{\label{qhist} The mass weighted global ionization fraction as
a function of redshift for the different reionization models.}
\end{figure}

In order to test the impact of X-ray contribution to reionization on
the kSZ effect we produce several different simulations using the
model above.  The model parameters which can be adjusted are $N_{\rm
ion}$, $C$, $M_{\rm min}$, $M_{\rm ion}$, $X_{\rm frac}$, and
$\lambda$.  We produce several simulations with different
parametrizations, but which have an optical depth to the surface of
last scattering, $\tau$, consistent with WMAP measurements
($\tau=0.088 \pm 0.015$) \cite{2011ApJS..192...18K}.  For simplicity,
we ignore the residual ionization from cosmological recombination
(prior to reionization).  From these models we can evaluate the effect
of a varying level of X-ray contribution to reionization.  The values of
the various parameters for each model are listed in Table
\ref{rmodels}. Due to the constraint on $\tau$, our models have fairly
similar global reionization histories.  These are shown in
Fig. \ref{qhist}.

\begin{table}
\caption{\label{rmodels} Parameters for the different reionization
models used.  For the minimum masses of dark matter halos which host
galaxies, the corresponding virial temperature (assumed to be
constant) is listed instead of the mass (see Eq. (\ref{tvir}) in the
text). }
\begin{tabular}{c c c c c c c c c}
Model & $N_{\rm ion}$ & $C$ & ${\rm T_{\rm vir}(M_{\rm min})}$ &${\rm T_{\rm vir}(M_{\rm ion})}$  & $X_{\rm frac}$ & $\lambda$ & $\tau$ \\
\hline
Fiducial  & 60  & 2  & $10^4$K & $10^5$K & 0 & - & 0.088 \\
Clump  & 355  & 20  & $10^4$K & $10^5$K & 0 & - & 0.088 \\
$M_{\rm min}$  & 115  & 2  & $5 \times 10^4$K & $10^5$K & 0 & - & 0.088 \\
$10\%$ X-ray  & 53  & 2  & $10^4$K & $10^5$K & 0.1 & 20 Mpc & 0.088 \\
$30\%$ X-ray  & 45  & 2  & $10^4$K & $10^5$K & 0.3 & 20 Mpc & 0.088 \\
$50\%$ X-ray  & 40  & 2  & $10^4$K & $10^5$K & 0.5 & 20 Mpc & 0.088 \\
\end{tabular}
\end{table}

\begin{figure*}
\includegraphics[width=3in]{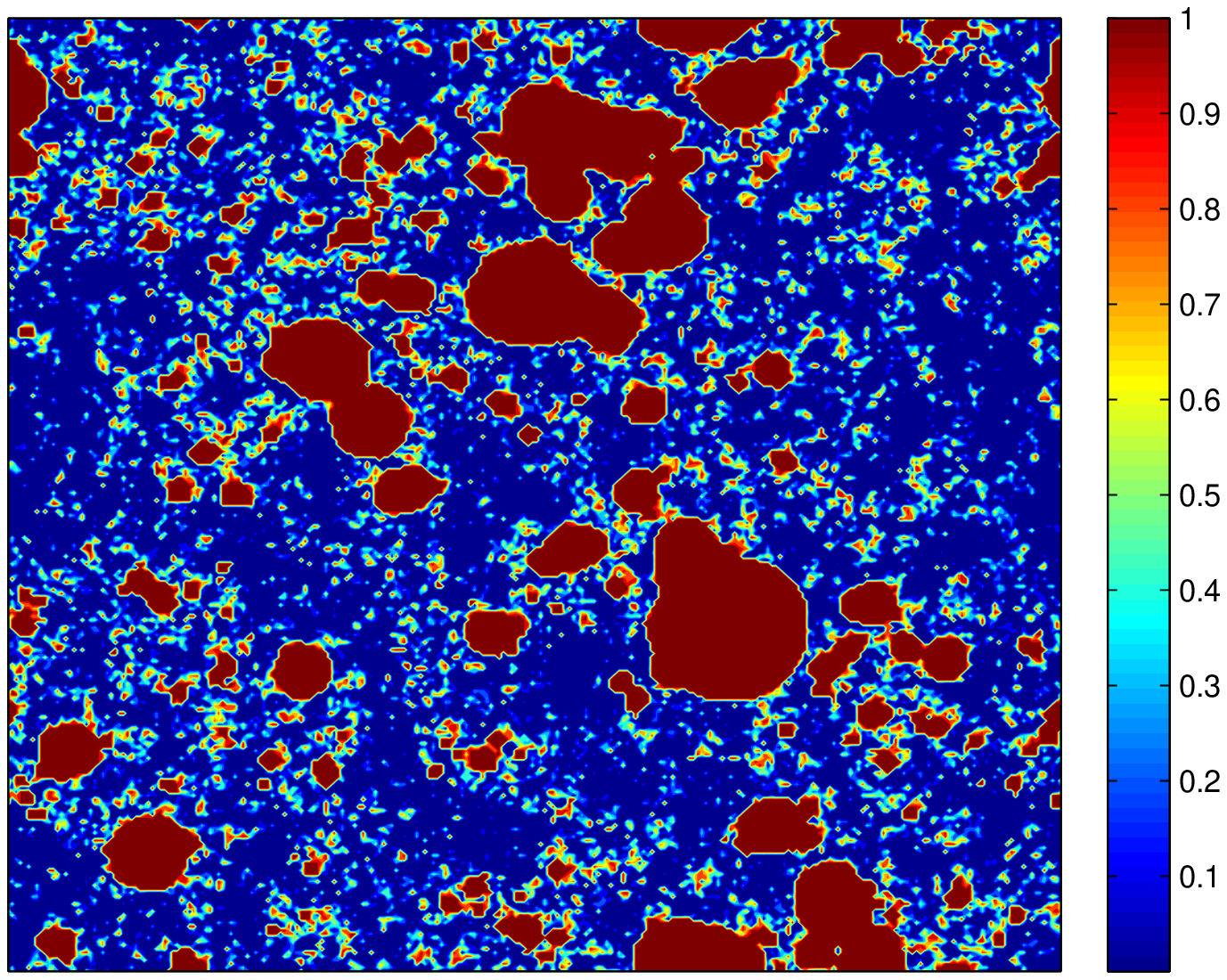}
\includegraphics[width=3in]{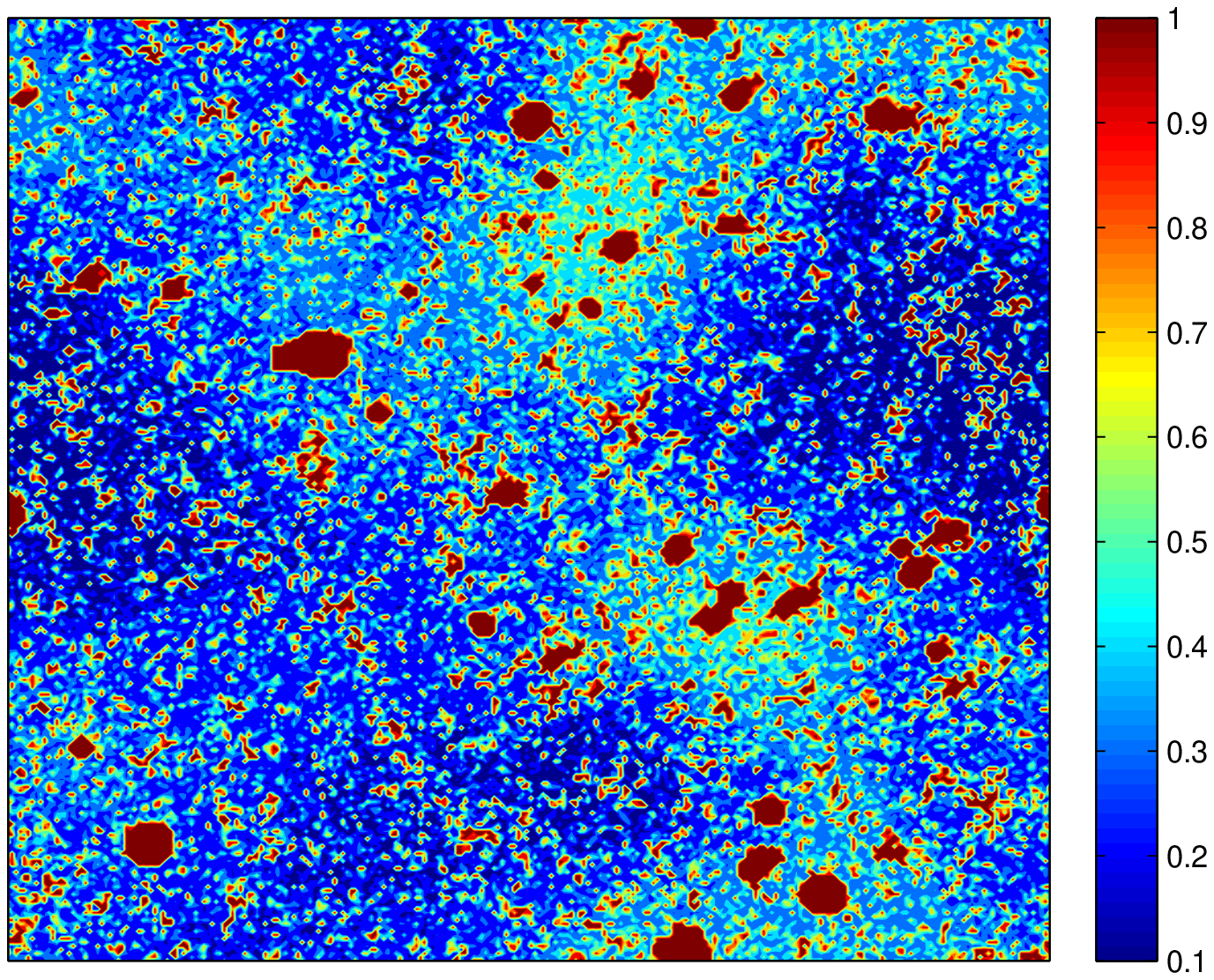}
\caption{\label{pic1} The ionization fraction through a slice of our
``fiducial'' (left panel) and ``50$\%$ X-ray'' (right panel) model at
$z=9$.  Note that the same underlying density field has been used for
both.  The slice corresponds to a $170 \times 170 \times 0.66 \: {\rm
Mpc}^3$ volume in the simulation.  The color bar gives the mass
weighted neutral fraction in each pixel. }
\end{figure*}

The main difference with X-rays is that their relatively long MFP
would cause some smooth ionization fraction throughout the IGM in regions
which are not inside bubbles.  This is illustrated clearly in
Figure \ref{pic1}.  We plot the ionization fraction
through a slice of our simulation at $z=9$ for the ``fiducial'' and
``$50\%$ X-ray'' models.  In both cases we have used the same
underlying density field and the global mass weighted ionization
fraction is roughly the same.  There is a striking difference in the
character of reionization between these two models.  The UV-only case
is dominated by completely ionized bubbles with surrounding neutral
regions.  The model with X-rays has smaller bubbles surrounded by a
partially ionized IGM.

In order to quantify the change in the kSZ signal due to X-rays, we
compute the angular power spectrum of CMB temperature fluctuations,
$C_l$.  We calculate the kSZ temperature fluctuations from the EoR
($z=6-20$) by integrating Eq. (\ref{dTintegral}) through our
simulations.  The resulting 2D temperature field is then Fourier
transformed to compute the power spectrum.  We include the kSZ signal
from lower redshift by calculating the power spectrum analytically and
adding it to the EoR result.  This is necessary because at low
redshifts non-linear effects become significant and they are not
accurately captured with the density fields we use from 21cmFAST.  We
calculate the analytic power spectrum using the method described in
\cite{2005ApJ...630..643M} (see their Eq. (4)).  To account for the
primordial temperature fluctuations, which dominate on large angular
scales, we add the lensed $C_l$'s calculated from CAMB \cite{camb}.
We make the assumption that the thermal Sunyaev-Zel'dovich effect,
with its distinctive spectral signature, will be removed completely.

\begin{figure}
\includegraphics{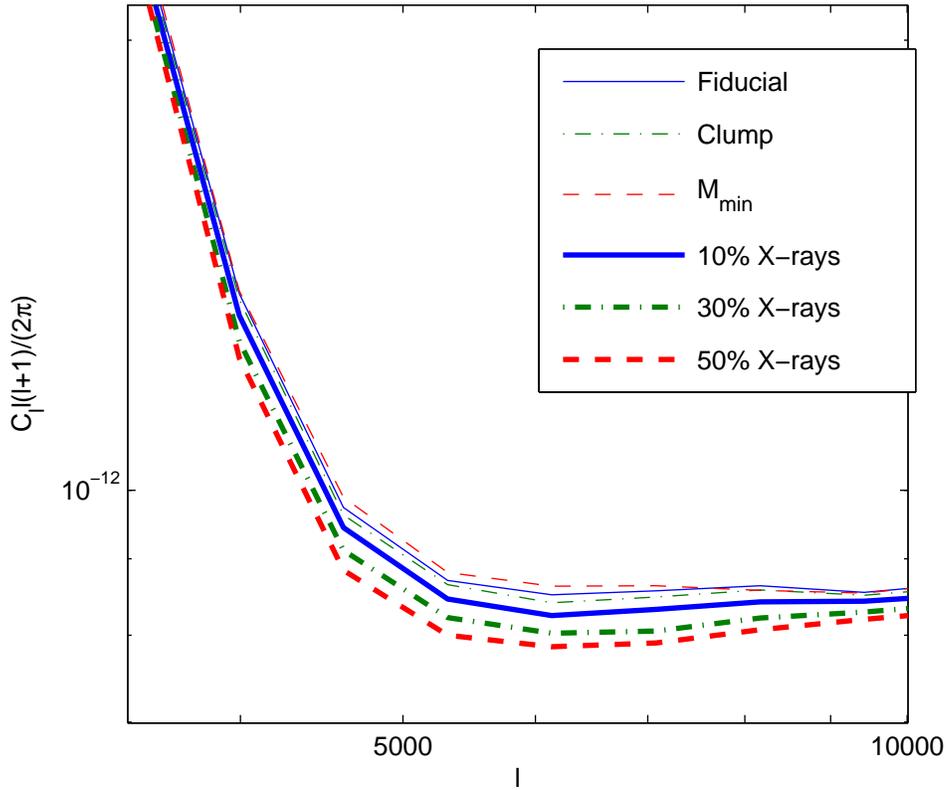}
\caption{\label{clplot} The angular power spectrum of temperature
fluctuations for the different reionization models tested.}
\end{figure}

All of the models tested were designed to have nearly identical
optical depth due to electron scattering.  Of course there will always
be some uncertainty in value of this quantify.  For example, Planck
measurements will constrain $\tau$ to within $\approx 0.005$
\cite{2006astro.ph..4069T}.  We find that this uncertainty should not
strongly affect our results.  As an example, for the ``fiducial''
model, we vary $N_{\rm ion}$ within the corresponding uncertainty in
$\tau$ and find that the power spectrum never goes as low as the
$10\%$ X-ray model for the $l$ values in Figure \ref{clplot}.

\begin{figure}
\includegraphics{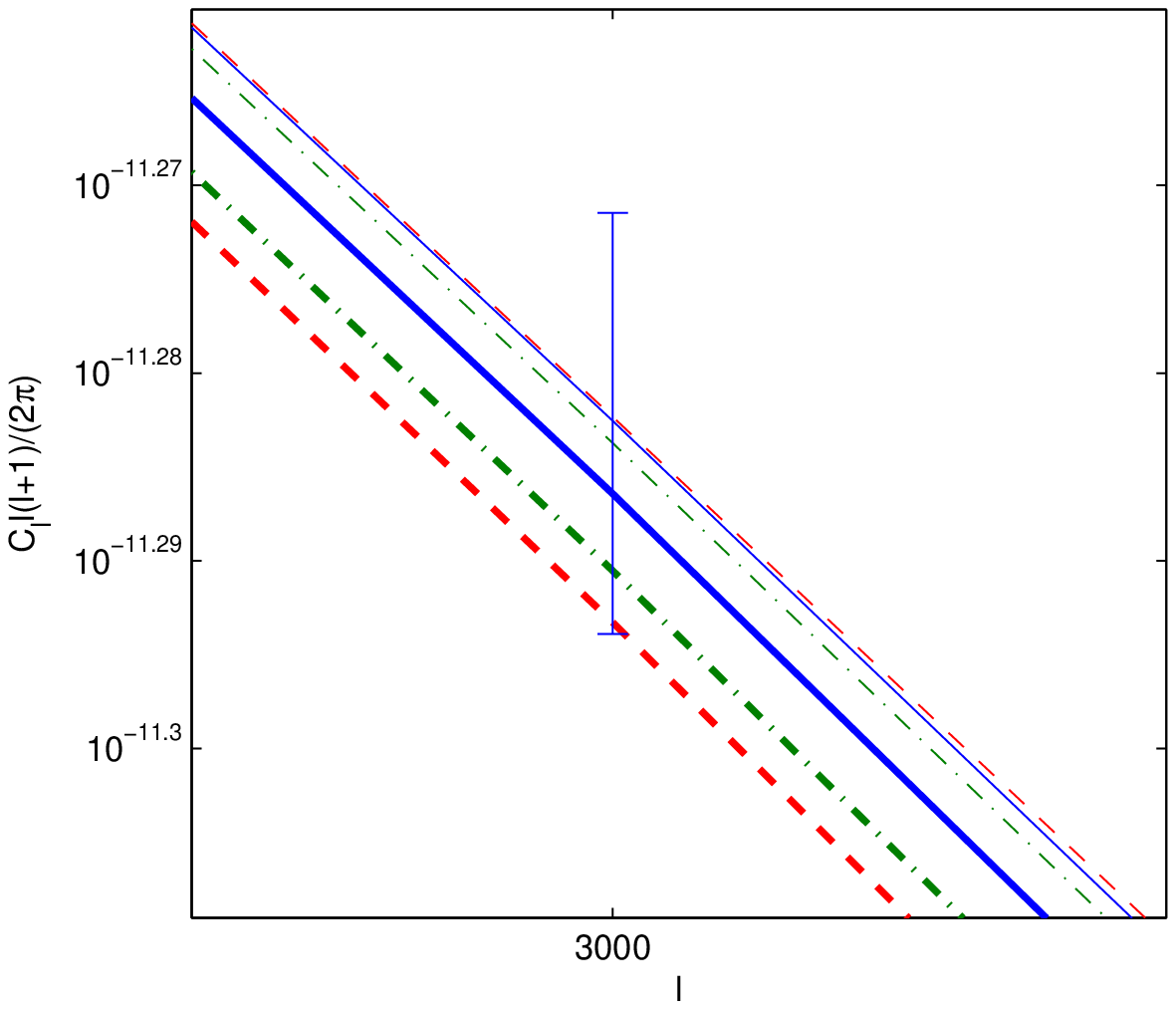}
\caption{\label{clplot2} The angular power spectrum of temperature
fluctuations at $l=3000$, where ACT and SPT will be sensitive to kSZ
signal.  The expected accuracy of SPT, $\Delta \left (
C_ll(1+l)/(2\pi) \right ) = 1 \mu \rm{K^2}/\bar{T}_{\rm CMB}^2$, is
shown as an error bar.  The various reionization models are shown with the
same lines as Fig. \ref{clplot}.}
\end{figure}

In addition to plotting the theoretical estimates, we also show the
error bars of SPT in Fig. \ref{clplot2} \cite{2004SPIE.5498...11R}.
SPT will only be sensitive to the kSZ signal at $l \approx 3000$ due
to the high amplitude of the primordial temperature fluctuations at
lower $l$ and dusty star-forming galaxies at higher $l$.  Detailed
estimates by the SPT team project an ultimate sensitivity of $\Delta
\left ( C_l l(1+l)/(2\pi) \right ) = 1 \mu \rm{K^2}/\bar{T}_{\rm
CMB}^2$.  ACT \cite{2003NewAR..47..939K} will be a couple times less sensitive (Matthew McQuinn,
private communication 2011).

\section{Discussion and Conclusions} 
We have calculated the effect of an X-ray contribution to reionization
on the fluctuations in the CMB from the kSZ effect.  This was
accomplished with semi-numerical simulations of reionization.  For the
case with no X-ray contribution, we tested how changes in the physical
parameters, including the number of ionizing photons produced, the
clumpiness of the IGM and the minimum mass of dark matter halos which
can host galaxies affect the kSZ signal.  This was then compared with
various levels of X-ray contribution to reionization parametrized by
the fraction of ionizing photons which are X-rays, $X_{\rm frac}$.

Our reionization model assumes that a constant number of ionizing
photons are produced for each baryon which is incorporated into galaxies.
We then calculate how many baryons have collapsed into dark matter
halos at a given redshift using the extended Press-Schechter
formalism.  UV photons have a MFP much smaller than the pixel scale in
our simulations while X-rays are assumed to uniformly ionize regions
with a fixed comoving MFP, for simplicity.  Varying the MFP between
$\lambda = 20$ Mpc and $\lambda = 50$ Mpc yields very similar results,
suggesting that our conclusions are robust.

Due to the comparatively long MFP of X-rays, they can partially ionize
the regions in between the ionized bubbles created by UV photons.  For
a fixed number of total ionizing photons this results in smaller
bubbles surrounded by partially ionized IGM.  This affects the kSZ
signal by reducing the ``patchy'' component of the fluctuations,
associated with the non-uniform topology of reionization.  In this way
it is possible to distinguish models which have a very similar global
reionization history but different X-ray contributions.

In particular, we find that an X-ray contribution greater than $X_{\rm
frac}=10\%$ produces a substantially lower power spectrum than our
UV-only models constrained to have equal total optical depth to
electron scattering.  This suggests that it may be possible to
constrain the X-ray contribution to reionization with future kSZ
observations.  Unfortunately, the sensitivity of SPT and ACT may be
insufficient to distinguish the models presented in this paper.
However, a non-detection of the kSZ signal from the EoR may suggest a
significant X-ray contribution.  We find that our X-ray models
push the kSZ power from reionization below the anticipated sensitivity
of SPT and ACT.  Because annihilation or decay of dark matter
particles could imprint a similar uniform ionization component, such a
non-detection could also be used to constrain the properties of dark
matter.

\section{Acknowledgments}
This work was supported in part by NSF grant AST-0907890 and NASA
grants NNA09DB30A and NNX08AL43G (for A.L.).

\bibliography{kSZ_xray}

\end{document}